# Visualizing and Optimizing Phase Matching in Nonlinear Guided-mode Resonators with the Green's Function Integral Method

Chengkang Liang,[1] Quanying Li,[1] Jiale Xu,[1] Pingqi Gao,[1] and Jiancan Yu[1,*]

[1]School of Materials, Sun Yat-sen University, Shenzhen, China, 518107

*yujc3@mail.sysu.edu.cn

**Abstract:** Efficient nonlinear frequency conversion in nanophotonics requires not only strong fundamental field but also precise phase matching among distributed nonlinear sources. Here, we develop the two-dimensional Green's function integral method (GFIM), which enables direct visualization and optimization of phase matching in nonlinear guided-mode resonators. Using GFIM phase analysis, we generalize the phase-matching factor (PMF) as a rigorous metric of spatial phase coherence in harmonic generation, revealing severe phase mismatch in conventional guide mode resonators. Guided by phase-matching profiles, we propose design strategies to improve the phase coherence, particularly by introducing a high-index waveguide layer that confines the fundamental field in the nonlinear material to regions where the harmonic Green's function varies slowly. This configuration achieves a PMF exceeding 0.91, approaching the ideal value of unity, and yields a record SHG efficiency of 26.7% at a low pump intensity of 2 kW/cm². These results establish the GFIM-based phase-matching visualization as an effective strategy for compact, high-performance nonlinear photonic devices.

Keywords: Second-harmonic generation, Nonlinear optics, Phase-matching factor, Guided-mode resonance, Phase analysis

## 1. Introduction

Nonlinear optical frequency conversion originates from the nonlinear dipole polarization of materials and underpins a wide range of applications in both classical and quantum optics, including integrated photonic circuits, ultrafast light sources, and quantum emitters [1–5]. Their efficiency (e.g., second-harmonic generation, SHG) critically depends on satisfying the phase-matching conditions, which ensure coherent buildup of the generated harmonic fields [6–8]. In bulk nonlinear media, phase matching is fundamentally limited by inherent wavevector mismatch caused by material dispersion [9,10]. To overcome this constraint, three principal strategies have been developed: exploiting crystal birefringence via anisotropic optical properties, angular tuning of the interacting waves to compensate dispersion, and quasi-phase matching via periodic sign modulation of the nonlinear susceptibility [11–15]. These approaches enable constructive interference of the generated harmonics, sustaining continuous energy transfer from the fundamental to the harmonic waves.

Micro/nanophotonic architectures have revolutionized nonlinear photonics by enhancing light–matter interactions through various high-Q resonant mechanisms [16–18], such as surface plasmon resonance (SPR), Fabry–Pérot resonance in distributed Bragg reflector (DBR) cavities, Mie resonance, and guided-mode resonance (GMR), thereby greatly improving SHG efficiency [19–23]. Although the recently emerging bound states in the continuum (BICs) have further boosted SHG efficiency owing to their theoretically infinite lifetime [24–28], the intrinsic phase mismatch dominated by structural rather than material dispersion in nanoscale systems has often been overlooked [29–31]. A series of practical paradigms have been proposed, including waveguide-based nonlinear systems where modal phase matching is evaluated using mode-overlap integrals [26,31–36], and quasi-phase-matching strategies extended to twist-engineered two-dimensional materials that compensate the phase by spatially tailoring $\chi^{(2)}$ tensor components [14,37]. However, localized phase mismatch in resonators are ubiquitous, while conventional wavevector-matching approaches are inherently macroscopic. Our recent development of a Green's function integral method (GFIM) allows for spatially resolved phase coherence analysis to quantitatively probe microscopic phase mismatch, achieving perfect phase matching [23]. Nevertheless, this case is limited to one-dimensional systems, whereas integrated nonlinear photonic platforms inherently require two- or even three-dimensional configurations.

Here, we extend the GFIM framework to two-dimensional GMR structures based on thin-film lithium niobate (LN), and generalize the concept of phase-matching factor (PMF). The PMF provides a compact quantitative measure of the spatial coherence of the constituent second-harmonic (c-SH) waves generated by microscopic nonlinear sources. The continuous destructive interference of c-SH radiation in traditional GMR structures results in lower PMF, preventing any substantial improvement in SHG efficiency. Guided by the GFIM-visualized c-SH phasor ellipses, we propose several designs to improve the PMF. In particular, we introduce a high-index waveguide layer to reshape and confine the fundamental field ($E_\omega$) to regions where the harmonic Green's function ($G_{2\omega}$) varies slowly. The resulting microscopic source, with intensity proportional to $G_{2\omega} E_\omega^2$, collectively generates a harmonic field whose constituent waves are nearly in-phase. Consequently, the optimized GMR structure achieves a PMF exceeding 0.91 and a peak SHG conversion efficiency of 26.7% under a low pumping intensity of 2 kW/cm$^2$. These findings demonstrate the capability of GFIM to capture complicated nonlinear interactions and provide a predictive, phase-engineered design framework for the rational design of compact, high-performance integrated nonlinear optical devices.

## 2. Results and discussion

### 2.1 Models and Methods.

Lithium niobate-on-insulator (LNOI) combines strong $\chi^{(2)}$ nonlinearity, large electro-optic coefficients, and low loss with tight optical confinement, forming an ideal platform for guided-mode resonance structures enabling efficient SHG [3,38,39]. In our design (**Fig. 1a**), an $x$-cut LN film is bonded to a silicon dioxide substrate, on top of which a periodic poly methyl methacrylate (PMMA) grating is patterned. The grating grooves are aligned parallel to the $c$-axis of LN (i.e., the $z$-axis). Detailed geometric parameters and material properties are provided in Section 1 of Supporting Information. The periodic PMMA grating enables fully electron-beam-lithography-fabricated devices, forming an all-dielectric metasurface [40]. This perturbation layer diffracts the incident light, coupling it into the waveguide modes and forming leaky modes that lie above the light cone in momentum space (Fig. 1b) [41,42].

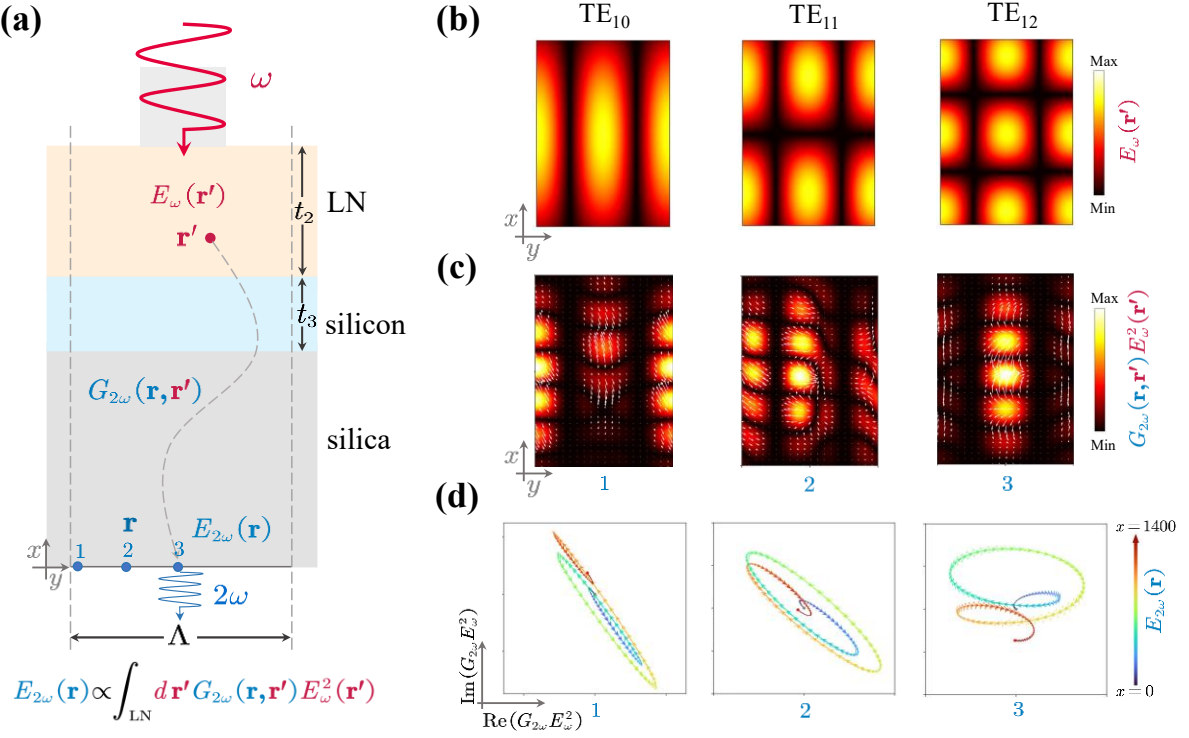

**Fig. 1. Visualizing SHG processes in a guided-mode resonator using the Green's-function integral method (GFIM).** (a) Schematic of the SHG process in a lithium niobate-based GMR structure. The structure consists of a periodic PMMA grating, a dual-layer waveguide composed of both a linear (non-$\chi^{(2)}$) and nonlinear (lithium niobate, $\chi^{(2)}$) layers, and a silica substrate. $t_2$ and $t_3$ denote the thicknesses of the two waveguide layers, while $\Lambda$ corresponds to the structural pitch (typically $\Lambda = 1$ μm). (b-d) Using $TE_{10}$ mode in conventional GMR ($t_2$ =1400 nm and $t_3 = 0$) as an example, the SHG process is visualized by GFIM. (b) The incident fundamental wave excites GMR modes ($E_\omega$) $TE_{10}$, $TE_{11}$, and $TE_{12}$ within the LN. (c) Spatial contribution of $G_{2\omega}E_\omega^2$ in LN for three representative radiation ports. Color denotes amplitude, and white arrows indicate the complex-plane vectors. (d) The dipole amplitude scales as $G_{2\omega}E_\omega^2$, and the total second harmonic field $E_{2\omega}$ arises from the coherent superposition of the c-SH waves from all microscopic nonlinear sources.

In nonlinear processes, phase matching arises from maintaining a uniform phase relationship among nonlinear dipole emitters induced by microscopic nonlinear polarization, enabling constructive interference of their radiated fields along the propagation direction. The GFIM models the total second-harmonic field by superposing the c-SH contributions from distributed nonlinear polarization sources, thereby naturally capturing the SHG process:[43]

$$\mathbf{E}(2\omega,\mathbf{r}) = 4\omega^2 \mu_0 \int d\mathbf{r}' \mathbf{G}(2\omega;\mathbf{r},\mathbf{r}') \cdot \mathbf{P}^{(2)}(2\omega;\mathbf{r}'), \tag{1}$$

where $\mathbf{P}^{(2)}(2\omega;\mathbf{r}')$ serves as the source term for SHG, and $\mathbf{G}(2\omega;\mathbf{r},\mathbf{r}')$ is the harmonic Green's function, which quantifies the outcoupling efficiency between the harmonic source at the position $\mathbf{r}'$ within LN and the observation site at $\mathbf{r}$ (e.g., Fig. 1c) [44]. To fully exploit the maximum nonlinear susceptibility of LN ($\chi^{(2)}_{33}$ = 30 pm/V), we employ a transverse-electric (TE) polarized fundamental pump, such that only the out-of-plane polarization contributes [7]

$$P^{(2)}_{2\omega}(x,y) = \varepsilon_0 \chi^{(2)}_{33} E^2_\omega(x,y). \tag{2}$$

And the c-SH field from an infinitesimal region within a nonlinear medium is proportional to the nonlinear polarization weighted by the Green's function:

$$dE_{2\omega}(x,y,x',y') \propto G_{2\omega}(x,y,x',y') E^2_\omega(x',y') dx' dy'. \tag{3}$$

Thus, the total SH field observed at the monitoring port, representing the coherent superposition of all c-SH contributions (Fig. 1d), can be expressed as:

$$E_{2\omega}(x,y) = \frac{4\omega^2 \chi^{(2)}_{33}}{c^2} \iint_{\text{LN}} dx' dy' \, G_{2\omega}(x,y,x',y') E^2_\omega(x',y'). \tag{4}$$

The complex-valued integrand $G_{2\omega} E^2_\omega$ is a phasor whose phase determines the interference behavior of the microscopic contributions. Efficient SHG requires the phase of $G_{2\omega} E^2_\omega$ to remain consistent over the nonlinear region, ensuring constructive interference; otherwise, phase variations lead to destructive interference and reduced conversion efficiency. To quantify the spatial coherence of c-SH waves at a given monitoring port $x$, we define the pointwise phase-matching factor (pPMF) at any point $y$ as:

$$\Phi(x,y) = \frac{\left| \iint_{\text{LN}} G_{2\omega}(x,y;x',y') E^2_\omega(x',y') dx' dy' \right|}{\iint_{\text{LN}} |G_{2\omega}(x,y;x',y') E^2_\omega(x',y')| dx' dy'}. \tag{5}$$

This dimensionless factor quantifies the degree of constructive interference among all c-SH contributions and is independent of the absolute pump amplitude, under the undepleted-pump approximation. A value $\Phi = 1$ corresponds to perfect phase matching, an ideal condition rarely achieved due to intrinsic spatial phase variations in the complex product $G_{2\omega} E_\omega^2$ throughout the nonlinear medium.

More complex than one-dimensional case, the SH intensity at the monitoring port can vary significantly across the port. To evaluate the degree of phase coherence over the entire port located at $x$, we introduce the overall phase-matching factor (oPMF, or simply PMF) as follows:

$$\Phi(x) = \frac{\left| \int_{-\Lambda/2}^{\Lambda/2} dy \iint_{\text{LN}} \mathbf{G}_{2\omega}(x,y;x',y') \mathbf{E}_\omega^2(x',y') dx'dy' \right|}{\int_{-\Lambda/2}^{\Lambda/2} dy \iint_{\text{LN}} |\mathbf{G}_{2\omega}(x,y;x',y') \mathbf{E}_\omega^2(x',y')| dx'dy'}, \qquad (6)$$

where $\Lambda$ denotes the grating period. This metric captures global constructive interference of all c-SH contributions across the port. By contrast, the arithmetic mean of the local phase-matching factor, $\overline{\Phi}(x) = \frac{1}{\Lambda} \int_{-\Lambda/2}^{\Lambda/2} \Phi(x,y) \, dy$, only reflects the average spatial coherence and neglects interference effects encoded in the numerator of Eq. (6). The pPMF provides a spatially resolved measure of phase coherence, while oPMF offers a more comprehensive evaluation for analyzing nonlinear interactions in complex nanostructured systems.

**2.2 Phase Mismatch in Conventional GMR Structure.**

We first examine how resonance modes evolve with nonlinear-layer thickness (1100–1400 nm) in a conventional GMR structure, motivated by the potential of resonant pumping to enhance nonlinear interactions. The transcendental dispersion relation (Eq. S11) for a four-layer planar waveguide is used to determine the photonic bands of both conventional and optimized GMR structures (Fig. S1). Eigenmode analysis using COMSOL confirms these as the first three TE waveguide modes, $TE_{10}$, $TE_{11}$, and $TE_{12}$, arising from first-order transverse diffraction (Fig. S2a). The linear spectrum provides the foundation for understanding enhanced light–matter interactions in GMR systems (Fig. S2b). Among them, the $TE_{10}$ mode exhibits the strongest field confinement and the highest quality factor, thus is the most favorable candidate for efficient nonlinear conversion (Fig. S2c).

In the nonlinear study, the SHG efficiencies of the TE$_{10}$ (Fig. S3a) and TE$_{12}$ (Fig. S3c) modes increase with the LN thickness. To quantify the correlation between fundamental fields $E_\omega$ and SHG efficiency, we define a field enhancement factor $\overline{E_{\text{norm}}^2}$

$$\overline{E_{\text{norm}}^2} = \frac{\int_{\text{LN}} |E_\omega(x,y)|^2 \, dV}{\int_{\text{LN}} |E_{\text{in}}(x,y)|^2 \, dV}, \tag{7}$$

where $E_{\text{in}}$ is the incident field amplitude. Using $\overline{E_{\text{norm}}^2}$ as a metric, we observe a strong, though not strictly linear, correlation between field enhancement and SHG efficiency for the TE$_{10}$ and TE$_{12}$ modes. In contrast, the TE$_{11}$ mode exhibits an anomalous decrease in SHG efficiency with increasing $\overline{E_{\text{norm}}^2}$ (Fig. S3b and **Fig. 2a**). These anomalies further suggest that SHG efficiency involves factors beyond field enhancement (Fig. 2b).

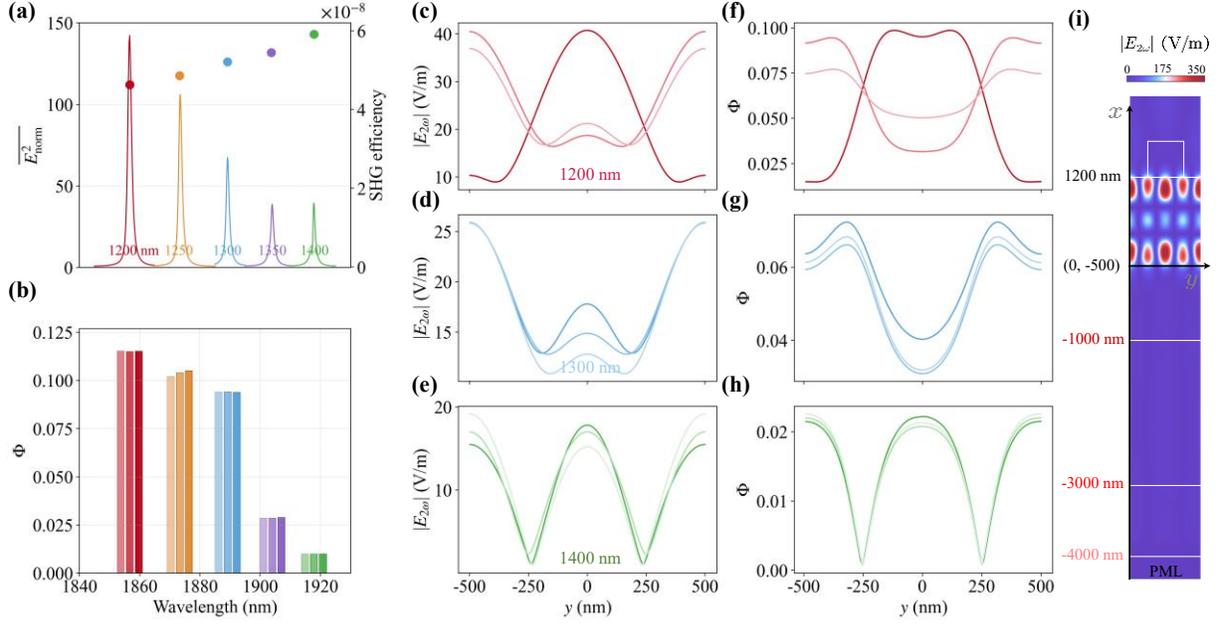

**Fig. 2. Phase-matching-factor analysis of the anomalous SHG in TE$_{11}$ mode.**
(a) Field enhancement factor $\overline{E_{\text{norm}}^2}$ (left axis, solid dots) and SHG efficiency (right axis, solid line) for the TE$_{11}$ mode pumping versus LN thickness ($t_2$ = 1200-1400 nm). (b) Overall phase-matching factor (PMF, $\Phi$) at resonance as a function of LN thickness. (c-e) Harmonic intensity profiles along the $y$-axis at different three radiation ports for $t_2$ = 1200 nm (b), 1300 nm (c) and 1400 nm (d). (f-h) Corresponding pointwise PMF for the same thicknesses. (i) Harmonic field distribution under TE$_{11}$ mode-pumping with an incident power density of 2 kW/cm² ($t_2$ = 1200 nm). Perfectly matched layers (PMLs) are applied above the excitation source and below the far-field port in simulations. Curves (columns) with varying transparency denote three different detection ports, with the farthest port corresponding to the most transparent color. Calculations across different detection ports show that pPMF (Eq. (5)) accurately captures any local phase matching, whereas oPMF (Eq. (6)) provides a robust measure of the overall phase-matching quality.

As the radiated harmonic field deviates from a plane wave (Fig. 2c-e), phase-matching evaluation in the two-dimensional model using the pointwise PMF becomes complex. To elucidate its spatial variation, we examine the pointwise PMF of the TE$_{11}$ mode along $y$ at three representative output ports (Fig. 2f-h). Taking the case of $t_2 = 1200$ nm as an example, the harmonic field distribution is shown in Fig. 2i. The pointwise values $\Phi(x_1,y)$ at $x_1 = -1000$ nm remain low (0.015 to 0.1), and exhibit strong variation across $y$, differing by up to a factor of 6. As the observation port moves farther away (i.e., $x \ll x_1$), the $x$-dependence of $\Phi(x,y)$ becomes less significant, while noticeable transverse variation persists. For other LN thicknesses, similar $x$ and $y$-dependence is observed, with $\Phi(x,y)$ consistently low. Despite minor deviations, the pointwise PMF $\Phi(x,y)$ captures local phase matching at any radiation point.

Using the overall PMF yields consistent values across all three ports for different LN thicknesses (Fig. 2b). indicating that the overall PMF effectively captures the global phase-matching behavior. This consistency likely stems from the conservation of second-harmonic power across output channels. The oPMF thus provides a convenient and reliable metric for evaluating overall phase-matching quality, and all subsequent analyses are based on it. This metric elucidates the anomalous dependence of SHG efficiency on the LN layer thickness $t_2$ for TE$_{11}$ mode (Fig. 2b): the PMF peaks at $t_2 = 1200$ nm despite the weakest field enhancement, whereas at $t_2 = 1400$ nm, stronger field enhancement coincides with a pronounced PMF reduction. The inverse PMF–field correlation shows that poor phase matching limits energy transfer from fundamental to harmonic waves. And the PMF values below 0.16 in all three modes strongly restrict SHG efficiency (Fig. S3d–f), highlighting the need to capture microscopic local phase mismatch.

We perform the detailed analysis on TE$_{10}$ resonant mode, which exhibits the strongest field amplitude but the lowest PMF $\Phi$ (Fig. S3). As shown in **Fig. 3a**, the normalized field intensity exceeded 40 ($\overline{E^2_{\text{norm}}} = 598$) at the resonance wavelength 2119.32 nm, in a 1400-nm-thick LN waveguide. Within the nonlinear region, TE$_{10}$ assumes two distinct phase states per period, differing by $\pi$, with two field nodes (Fig. 3b). This stems from the standing-wave resonance condition, which locks the lateral field phase into two discrete states and suppresses continuous spatial evolution. Consequently, the phase of $E^2_\omega$ remains constant, and the overall phase of the product $G_{2\omega} E^2_\omega$ is determined solely by the Green's function $G_{2\omega}$. This behavior is consistent with the resonance-induced phase-locking observed previously in cavity-integrated one-dimensional structures [23].

As shown in Fig. 3c-d, the harmonic Green's function to the port center ($y = 0$) oscillates nearly periodically in both amplitude and phase along the propagation direction, forming multiple cycles within the LN waveguide. Since the Green's function characterizes the point-to-point response, the product $G_{2\omega} E_\omega^2$ reflects the nonlinear contribution at a specific observation point. The pronounced variation of $G_{2\omega} E_\omega^2$ maps across different locations (Fig. S5) indicates that microscopic phase mismatch at any single point can be reliably inferred. Meanwhile, to capture the overall contribution across the entire port, we evaluate the line integral

$$\mathcal{I}[G_{2\omega} E_\omega^2] = \int_{\text{Port line}} G_{2\omega}(x,y) E_\omega^2(x,y) \, dy, \tag{8}$$

and visualize its magnitude and complex phasor in Fig. 3e. This integrated phasor coherently accounts for the collective radiation toward all port positions. Due to the steep spatial phase variation of $G_{2\omega}$, adjacent maxima in $|\mathcal{I}[G_{2\omega} E_\omega^2]|$ may possess nearly opposite phases ($\approx \pi$ shift) in $\mathcal{I}[G_{2\omega} E_\omega^2]$, leading to destructive interference among c-SH waves (Fig. S6).

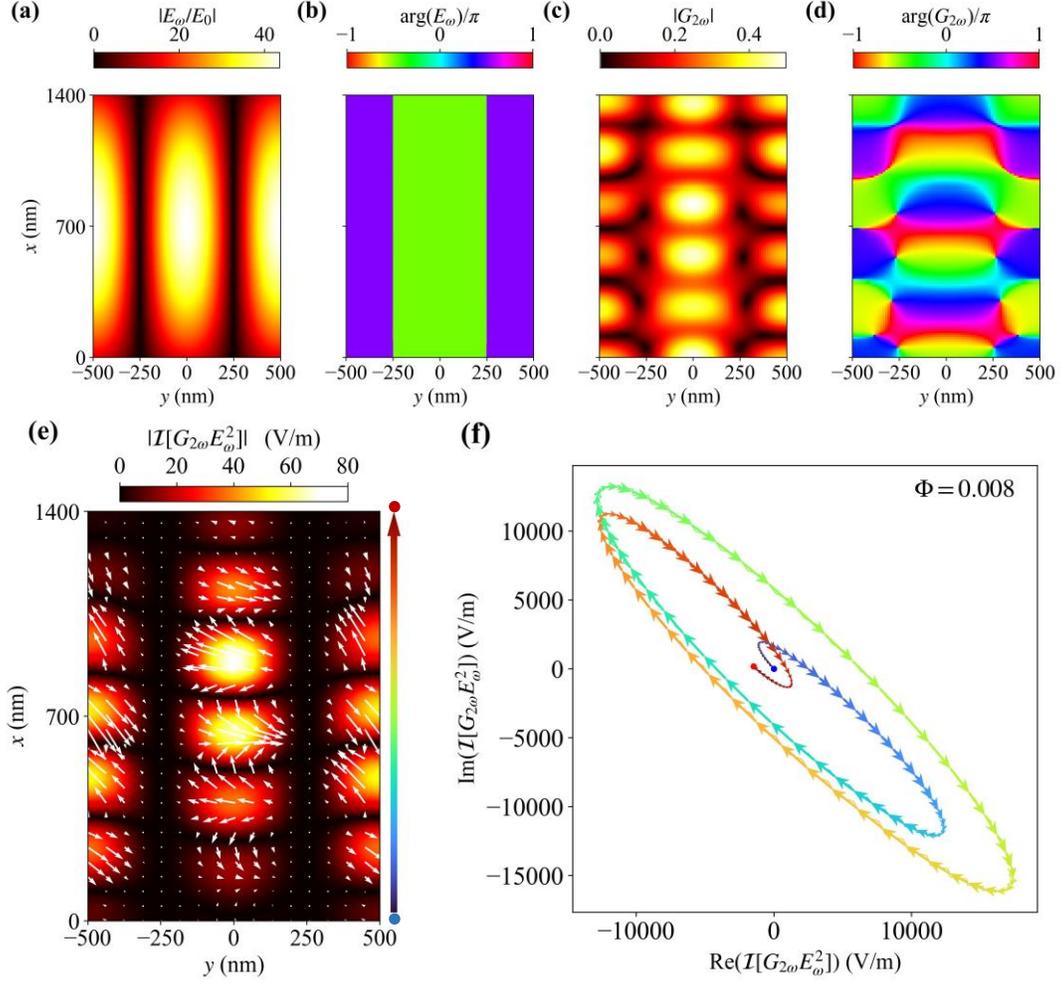

**Fig. 3. Microscopic phase-mismatch analysis using the GFIM.** (a, b) Amplitude (a) and phase (b) distributions of the fundamental TE$_{10}$ mode at $t_2 = 1400$ nm. A phase discontinuity by $\pi$ occurs across the field node along the $y$-axis, leading to a uniform phase distribution of $E_\omega^2$. (c, d) Amplitude (c) and phase (d) of the Green's function at the center of the detecting port at $x = $ -4000 nm. (e) Distribution of the product $G_{2\omega}E_\omega^2$ after integration along the $y$-direction at the port. White arrows denote the vector representation of $\mathcal{I}[G_{2\omega}E_\omega^2]$. (f) Sequential head-to-tail connection of these vectors along the $x$-axis (from blue to red) illustrates the superposition process of $\mathcal{I}[G_{2\omega}E_\omega^2]$. Each arrow represents the cumulative effect over a layer. Continuous destructive interference of the $\mathcal{I}[G_{2\omega}E_\omega^2]$ yields the elliptical trajectory, resulting in an overall PMF of only 0.008.

We further visualize the vector summation of $\mathcal{I}[G_{2\omega}E_\omega^2]$ across the entire LN region. This operation represents the phasor summation of local nonlinear polarization sources, with each phasor's argument given by the $\mathcal{I}[G_{2\omega}E_\omega^2]$ phase. In one-dimensional models, phasors sum along a single direction, whereas in two-dimensional geometry, column-wise integration produces highly disordered phasor accumulation patterns (Fig. S7), and row-wise summation yields a more organized, approximately elliptical trajectory in the complex plane (Fig. 3f). Although certain segments along the major axis exhibit large local magnitudes, they are largely canceled by phase-opposed contributions elsewhere along the ellipse. Consequently, the net harmonic amplitude remains small, with an overall PMF of only 0.008. These results suggest that harnessing locally constructive regions could provide an effective route to boost nonlinear conversion in otherwise phase-mismatched photonic systems.

**2.3 Design Optimization Guided by Phase Analysis**

Although phase matching in nanostructures is generally less pronounced than field enhancement, it remains a critical factor governing SHG efficiency. Beyond mode engineering, approaches such as wavefront shaping and structured-light illumination have been demonstrated to tailor the phase of the fundamental field and thereby enhance phase matching [26,45,46]. Leveraging the simplicity of angle tuning, we employ oblique incidence to reshape the fundamental field. As shown in Fig. S8, despite a noticeable decrease in field enhancement, the PMF markedly increases from 0.008 to 0.13 at an incidence angle of 41.95°, resulting in nearly unchanged SHG efficiency. Nevertheless, the near-periodic radiative nature of the Green's function in conventional GMR structures ultimately constrains performance. Under non-resonant pumping, mode tuning can, in principle, yield better phase matching (e.g., a PMF of 0.82 for a fundamental wavelength of 2000 nm), but the reduced field strength suppresses SHG efficiency; we therefore do not pursue this scenario further.

To extend the analysis, we apply the GFIM framework to reported nonlinear photonic structures, exemplified by quasi-BIC modes in the period-doubled GMR configurations [47]. The phasor summation of $\mathcal{I}[G_{2\omega}E_\omega^2]$ forms an approximate half-ring pattern (Fig. S9), analogous to half of the elliptical trajectory shown in Fig. 3f, leading to a substantial increase in PMF up to 0.28. This improvement primarily arises from restricting the Green's function variation to within one period via reduced LN thickness, combined with resonance-induced field enhancement.

Motivated by the partial phasor summation along the elliptical trajectory that enhances the PMF (Fig. 3f and Fig. S9), we optimize phase matching by replacing the substrate-adjacent region with a refractive-index-matched, non-$\chi^{(2)}$-active material. In **Fig. 4a**, by preserving the $TE_{10}$ profile (keeping $t_2 + t_3 = 1400$ nm), this modification suppresses destructive interference by removing opposite-phase contributions near the substrate. We track the $TE_{10}$ mode field enhancement factor $\overline{E_{\text{norm}}^2}$, SHG efficiency, and overall PMF as functions of non-$\chi^{(2)}$ thickness $t_3$ (Fig. 4b-d). The field enhancement exhibits a nonmonotonic trend, rising slightly before decreasing, whereas the PMF oscillates and approaches unity as $t_3$ approaches the total thickness. The SHG efficiency reflects a balance among the nonlinear interaction volume, the local field enhancement, and the phase-matching conditions. At $t_3 = 590$ nm, the SHG efficiency reaches its maximum of $6.8 \times 10^{-6}$ (40-fold enhancement), corresponding to an overall PMF of 0.21 (26-fold increase) and a nearly unchanged $\overline{E_{\text{norm}}^2}$ (1.05-fold). In ultrathin nonlinear layers, SHG efficiency drops below $10^{-9}$ due to the vanishing interaction volume, even though the PMF approaches its theoretical maximum. This explains why phase mismatch is often neglected in atomically thin materials. By redistributing nonlinear material to remove phase-destructive regions enables field confinement and phase alignment, substantially enhancing SHG efficiency.

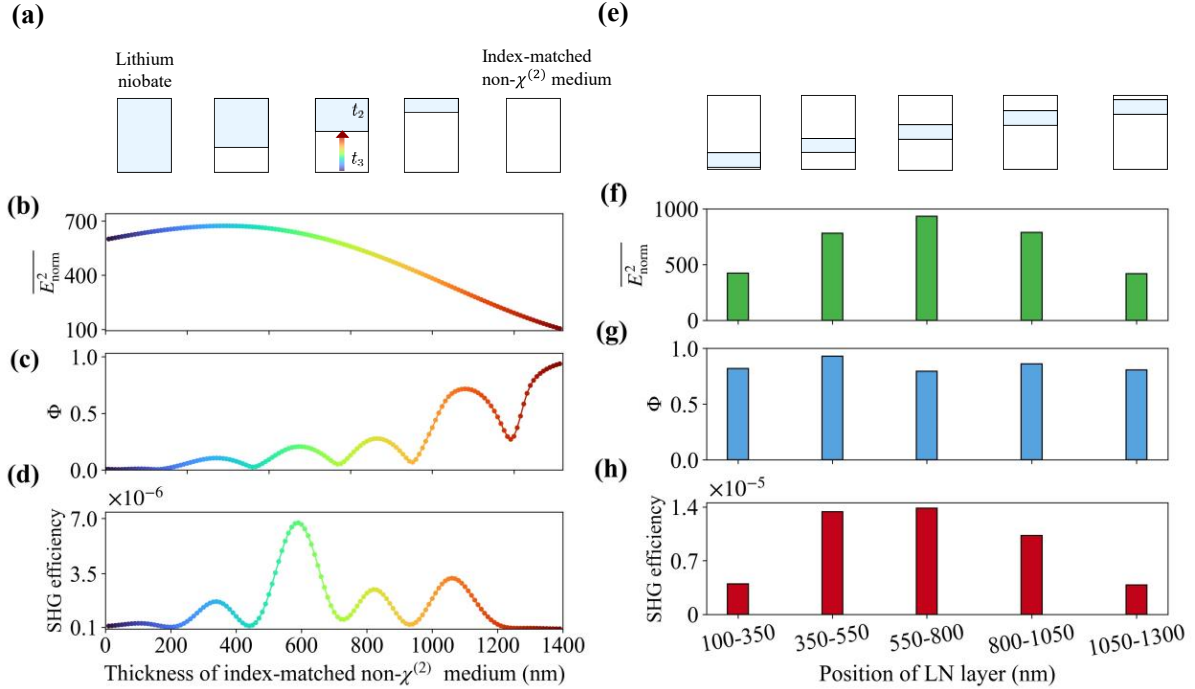

**Fig. 4. Two strategies for optimizing the phase-matching factor based on superposition trajectory.** (a) Schematic of a structure where the nonlinear layer is partially replaced by refractive-index-matched non-$\chi^{(2)}$ dielectric material. (b-d) Field enhancement factor (b), PMF (c) and SHG efficiency (d) as a function of non-$\chi^{(2)}$ material thickness. (e) Schematic of localized LN incorporation to support constructive interference. (f-h) Field enhancement factor (f), PMF (g), and SHG efficiency (h) for various designs. PMF remains consistently high ($> 0.75$) over a broad parameter range, while SHG efficiency follows the trend of field enhancement owing to favorable phase matching.

We further selectively retain the inner 200–250 nm of the LN layer, corresponding to the elliptical phasor arc that contributes predominantly constructive interference in the complex plane (Fig. 4e). The remaining region is replaced with a refractive-index-matched, non-$\chi^{(2)}$-active material to preserve the fundamental field. In this configuration, both a relatively high PMF ($\Phi > 0.75$) and a significantly enhanced SHG efficiency of $10^{-5}$ are achieved (Fig. 4f-h). Remarkably, reducing the nonlinear layer to only 200–250 nm doubles the SHG efficiency compared with the previous 600 nm optimized design (Fig. 4d). These results show that the phasor-trajectory metric reliably identify spatial regions that contribute constructively to SHG, enabling targeted placement of nonlinear polarization sources for improved performance. Nevertheless, the absolute SHG conversion efficiency remains moderate, as the local field enhancement is not yet fully optimized.

**2.4 Silicon–Lithium Niobate Dual Waveguide Design for Simultaneous Field and Phase Matching Optimization.**

To further enhance the field effect, we replace the previously index-matched, non-$\chi^{(2)}$ layer with the high-refractive-index silicon waveguide ($t_3 \neq 0$ in Fig. 1). **Fig. 5a** compares the field enhancement factor $\overline{E_{\text{norm}}^2}$, phase-matching factor $\Phi$, and SHG efficiency for the dual-waveguide configurations. The normalized field amplitude in LN increases to 380, corresponding $\overline{E_{\text{norm}}^2}$ reaches $10^4$—a 20.6-fold improvement over the silicon-free design. This enhancement arises from strong $TE_{10}$ confinement in silicon layer (Fig. 5b), with its evanescent tail efficiently coupling into the nonlinear LN. Importantly, the phase distribution of the fundamental field remains unchanged under resonant condition (Fig. 5c), so the phase of c-SH waves continues to be governed by the Green's function (Fig. 5d, e). Since the enhanced field is now concentrated in a thinner LN section where $G_{2\omega}$ exhibits a slowly varying spatial phase, the phasor values of $\mathcal{I}[G_{2\omega}E_\omega^2]$ (indicated by white arrows) become nearly aligned in regions of large magnitude (Fig. 5f and Fig. S10). Row-wise accumulation of phasors across the LN layer affords a smooth trajectory with a slightly curled tail in the complex plane (Fig. 5g), indicating enhanced constructive interference compared with the case in Fig. 3f. Under these conditions, the PMF reaches 0.63, and the SHG efficiency rises to $6.5 \times 10^{-3}$, corresponding to a $3.82 \times 10^4$-fold improvement over the original structure, and a 465-fold improvement relative to the design with a partially preserved LN layer in Fig. 4h (middle panel).

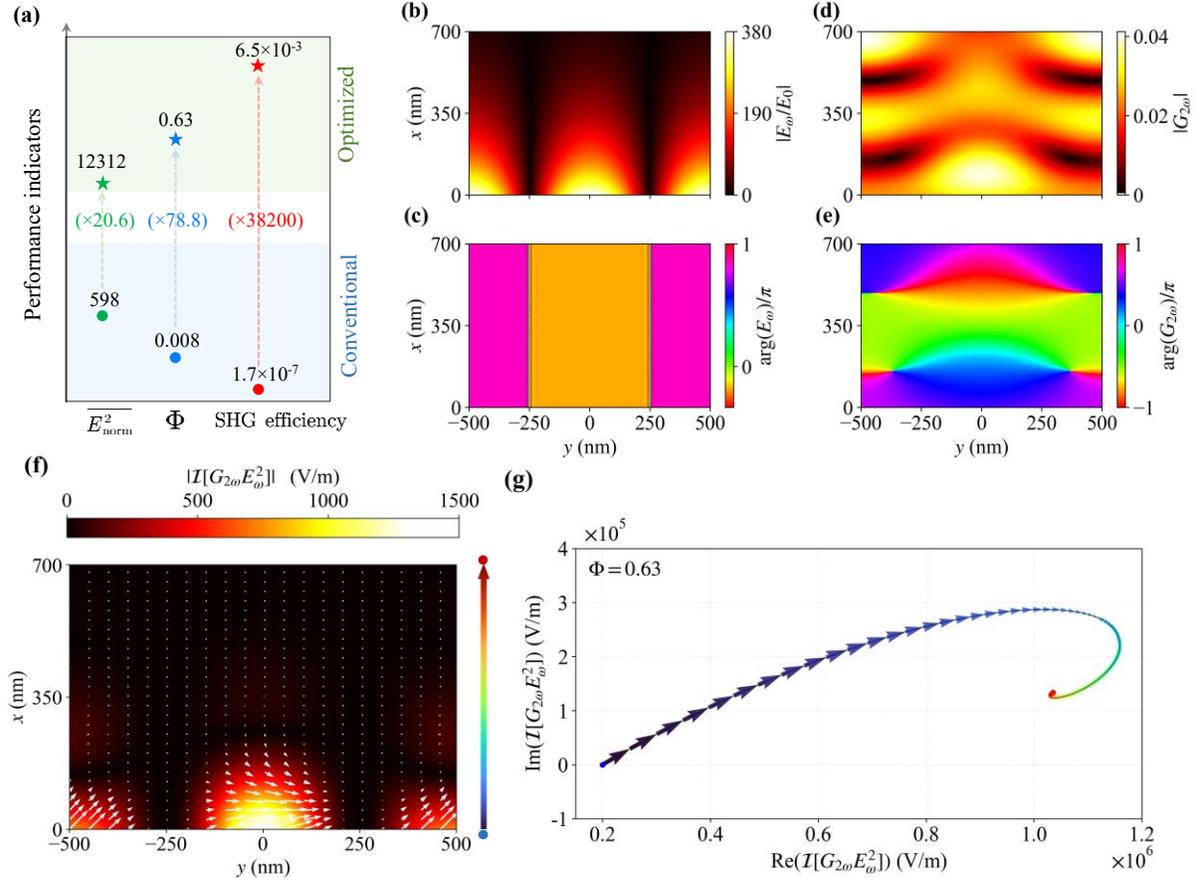

**Fig. 5. Phase-matching analysis of the optimized dual-waveguide GMR structure**. (a) Comparison of key performance metrics between the optimized ($t_2 = 400$ nm and $t_3 = 700$ nm) and the conventional ($t_2 = 1400$ nm and $t_3 = 0$) GMR structure. Values in parentheses denote the enhancement factors relative to the conventional configuration. (b-c) Amplitude (b) and phase (c) distributions of the fundamental $TE_{10}$ mode in the LN layer. (d-e) Amplitude (d) and phase (e) of the Green's function to the center of the output port. (f) Amplitude distribution of $\mathcal{I}[G_{2\omega}E_\omega^2]$, primarily concentrated in regions where the phase of Green's function varies slowly. (g) Vector superposition of the phasor $\mathcal{I}[G_{2\omega}E_\omega^2]$, revealing a favorable phase matching picture.

To evaluate the robustness of phase matching in the optimized GMR configuration, we examine how variations in LN and silicon waveguide layer thicknesses affect SHG efficiency (Fig. S11). Under favorable phase-matching conditions, increasing either layer thickness enhances SHG efficiency through stronger field confinement. In particular, a modest increase in LN thickness boosts the SHG efficiency up to 2.75%. Beyond layer thickness, the fundamental field in the LN can be further amplified by modulating the duty cycle of the PMMA gratings. Rigorous coupled-wave analysis (RCWA)[48] shows that the resonance peak near 2883.5 nm as the PMMA duty cycle increased from 0.1 to 0.9, while the full-width at half-maximum narrows at both extremes, indicating a higher quality factor (**Fig. 6a**). Correspondingly, at a duty cycle of 0.9, the maximum field enhancement factor and SHG efficiency increase by factors of 6 and 40, respectively (Fig. 6b). At a low pump intensity of 2 kW/cm², the optimized structure reaches 26.7% SHG efficiency, exceeding previous reports by over an order of magnitude (Table 1). This value is computed from a coupled model that accounts for pump depletion (Fig. S12). Under robust phase matching, optimizing field enhancement enables highly efficient frequency conversion even at low pump intensities.

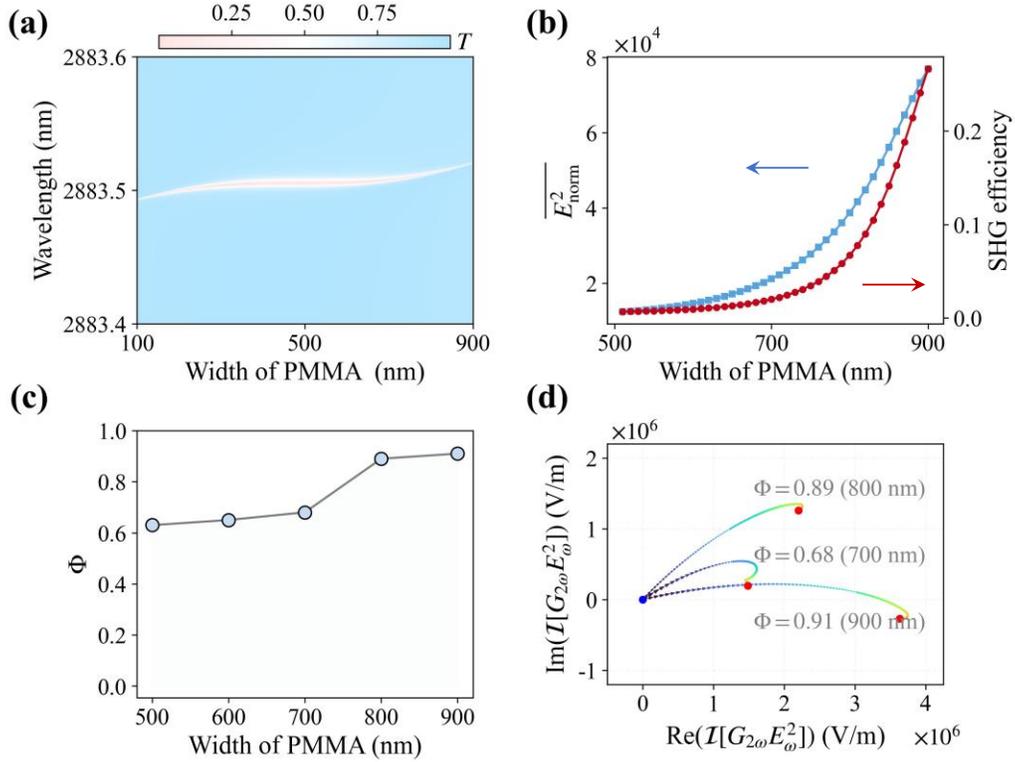

**Fig. 6. Optimization of SHG efficiency via optimized field enhancement and phase-matching.** GFIM analysis with varying PMMA width shows that the optimized phase-matching factor enhances the effectiveness of the field enhancement, leading to a breakthrough in SHG efficiency. (a) Transmission spectra versus fundamental wavelength and PMMA width. (b) Field enhancement factor $\overline{E_{\text{norm}}^2}$ (left axis, blue) and SHG efficiency (right axis, red) as functions of PMMA width. (c) Phase matching factor varies with the width of PMMA. (d) Vector superposition of $\mathcal{I}[G_{2\omega}E_\omega^2]$ for PMMA widths of 700 (blue), 800 (purple), and 900 nm (green), showing that higher field enhancement increases the fraction of constructive interference of the subharmonic generated by microscopic nonlinear sources, thereby improving the phase-matching factor.

We display the process of microscopic phase matching. At a duty cycle of 0.9, the harmonic field evolves toward a plane-wave profile away from the grating, causing the pointwise PMF at the output port to converge to the overall PMF (Fig. S13). The PMF increases from 0.63 to 0.91 (Fig. 6c), approaching near-perfect phase matching. It confirms that the observed SHG enhancement cannot be attributed solely to the fundamental-field amplification. Fig. 6d further illustrates that the improved phase matching shortens the curled tail of phasor trajectory of $\mathcal{I}[G_{2\omega}E_\omega^2]$, yielding favorable constructive interference. Field enhancement sets the theoretical upper bound of SHG efficiency, whereas phase matching dictates the practically achievable limit, serving as a decisive lever for engineering high-performance nonlinear photonic structures.

**Table 1.** Comparison of SHG efficiency enhancement in different LN based structures.

| Structure | Input power (MW/cm$^2$) | SHG efficiency | Reported year | Ref. |
|---|---|---|---|---|
| LN waveguide GMR | $1.3\times10^3$ | $8.1\times10^{-5}$ | 2021 | [17] |
| LN on Bragg reflector | $1.3\times10^3$ | $6.0\times10^{-3}$ | 2024 | [20] |
| Two-layer LN metasurface | $5.3\times10^3$ | $1.4\times10^{-4}$ | 2021 | [39] |
| LN waveguide GMR | $4.0\times10^{-1}$ | $5.0\times10^{-2}$ | 2022 | [28] |
| LN metasurface | $3.3\times10^{-3}$ | $4.9\times10^{-3}$ | 2021 | [24] |
| LN photonic crystal slab | $2.0\times10^{-3}$ | $1.0\times10^{-2}$ | 2023 | [18] |
| LN photonic crystal slab | $2.0\times10^{-3}$ | $3.7\times10^{-2}$ | 2023 | [16] |
| LN waveguide GMR | $5.0\times10^4$ | $1.36\times10^{-3}$ | 2025 | [40] |
| LN photonic crystal slab | $2.0\times10^{-3}$ | $2.26\times10^{-2}$ | 2025 | [38] |
| LN waveguide GMR | $1.0\times10^0$ | $2.8\times10^{-2}$ | 2023 | [27] |
| LN and silicon dual-waveguide GMR | $2.0\times10^{-3}$ | $2.67\times10^{-1}$ | | **Our work** |

### 3. Conclusions and outlook

In summary, we extend the Green's function integral method to visualize and optimize phase matching in LN-based GMR structures. We reveal that the counterintuitive decrease of SHG efficiency with increasing field enhancement in conventional GMRs arises from severe phase mismatch. Guided by the phasor-summation picture provided by GFIM, we introduce a high-index waveguide layer that redistributes the fundamental field into regions where the harmonic

Green's function varies slowly, achieving a PMF exceeding 0.91 and a high SHG conversion efficiency of 26.7% under a low pump intensity of 2 kW/cm².

These findings establish the phase-matching factor as a decisive design metric that complements conventional field-enhancement considerations, enabling predictive and phase-engineered strategies for various nonlinear nanophotonic devices. Looking forward, the GFIM framework can be generalized to fully three-dimensional architectures, higher-order and difference-frequency generation, as well as non-Hermitian and topological photonic platforms where coherent superposition is essential. Moreover, since the GFIM inherently captures information on coherent interaction, it may also provide a valuable tool for quantum photonic platforms where coherence must be precisely controlled or intentionally suppressed.

## 4. Numerical simulation method

We validated the Green's function method via finite-element simulations in COMSOL. Under the undepleted-pump approximation, the fundamental and harmonic fields are solved sequentially, with the nonlinear polarization driving the SH field. For pump depletion, the mutual coupling between fundamental and harmonic fields is fully accounted for via self-consistent solution of the coupled-wave equations. The total SHG power is obtained by integrating the time-averaged Poynting vector of the radiated SH field at output port. To compute the Green's function describing radiation from a unit-area polarization source ($J_z = -j\omega P_z$), an out-of-plane current is swept across the LN region. This procedure directly maps the system's linear response, establishing a clear link between the fundamental field, harmonic Green's function, and generated SH output.

**Author information**

**Corresponding Authors**

*Email: yujc3@mail.sysu.edu.cn

**ORCID**

Jiancan Yu: 0000-0001-9723-6113

**Author Contributions**

J.Y. conceived the original concept, and designed the research framework. J.Y. and C.L. prepared the manuscript draft and completed the manuscript. C. L. performed numerical simulations. J.Y., C.L., and P.G. contributed to critical revisions and scientific refinement of the manuscript. All authors reviewed and approved the final version of the manuscript.

**Code and Data Availability**

The code and data that support the findings of this study are available from the corresponding author upon reasonable request.

**Acknowledgments**

The authors acknowledge the financial support provided by National Natural Science Foundation of China (No. 62175264).

# Supporting Information: Visualizing and Optimizing Phase Matching in Nonlinear Guided-mode Resonators with the Green's Function Integral Method

Chengkang Liang,[1] Quanying Li,[1] Jiale Xu,[1] Pingqi Gao,[1] and Jiancan Yu[1,*]

1School of Materials, Sun Yat-sen University, Shenzhen, China, 518107

*yujc3@mail.sysu.edu.cn

## 1. Dispersion relationship of double-layer waveguide model

To exploit the guided-mode resonance (GMR) for enhancing the fundamental field and thereby achieving higher second-harmonic generation (SHG) efficiency, we first analyze the dispersion relation of the transverse electric (TE) mode in a more comprehensive four-layer planar waveguide structure, as shown in Fig. S1. Treating periodic PMMA grating as perturbation ($n_1 = 1.47$), the structure consists of air ($n_0$), lithium niobate (LN) slab ($n_2 = 2.2$), Si film ($n_3 = 3.47$) and semi-infinite thickness silica substrate ($n_4 = 1.43$). Under the $z$-polarized light incidence, the planar transverse electric wave appears in the form of $\mathbf{E}_j(x,y) = \hat{\mathbf{z}} E_{j,z}(x) e^{i\beta y}$ ($j = 0, 2, 3, 4$), and $\beta$ represents the propagation constant[1]

$$\beta = \frac{2\pi n_{\text{eff}}}{\lambda} = \frac{2\pi}{\lambda}\sin\theta \pm \frac{2\pi m}{\Lambda}, \quad (m = 0, 1, 2...) \tag{S1}$$

where $\theta$ is the incident angle, $\Lambda$ is the virtual pitch, and $m$ is the diffraction order. Electric field mode in each layer of material satisfies the wave equation

$$\frac{\partial^2 E_{j,z}}{\partial x^2} + (k_0^2 n_j^2 - \beta^2) E_{j,z} = 0, \quad (j = 0, 2, 3, 4), \tag{S2}$$

where $k_0 = \omega/c$, $\omega$ is the angular frequency and $c$ is the light speed in vacuum. Specifically, the distribution of electric field in each layer of the four-layer waveguide is described as:

$$E_z(x) = \begin{cases} A e^{-\beta_0(x-t_2)}, & (t_2 < x < +\infty) \\ B\cos(\beta_2 x) + C\sin(\beta_2 x), & (0 < x < t_2) \\ D\cos(\beta_3 x) + E\sin(\beta_3 x), & (-t_3 < x < 0) \\ F e^{\beta_4(x+t_3)}, & (-\infty < x < -t_3) \end{cases} \tag{S3}$$

where $A - F$ represent their amplitudes. Substituting Eq. (S3) into Eq. (S2), we get the detailed wavevector components:

$$\begin{cases} \beta_0 = (\beta^2 - n_0^2 k_0^2)^{1/2}, \\ \beta_2 = (n_2^2 k_0^2 - \beta^2)^{1/2}, \\ \beta_3 = (n_3^2 k_0^2 - \beta^2)^{1/2}, \\ \beta_4 = (\beta^2 - n_4^2 k_0^2)^{1/2}. \end{cases} \tag{S4}$$

By applying the continuity condition of electric field boundary at $x = t_2, 0, -t_3$, we obtain:

$$\begin{cases} A = B\cos(\beta_2 t_2) + C\sin(\beta_2 t_2), \\ B = D, \\ F = D\cos(\beta_3 t_3) - E\sin(\beta_3 t_3). \end{cases} \tag{S5}$$

Therefore, we can rewrite the expression for the electric fields:

$$E_z(x) = \begin{cases} [D\cos(\beta_2 t_2) + C\sin(\beta_2 t_2)]e^{-\beta_0(x-t_2)}, & (t_2 < x < +\infty) \\ D\cos(\beta_2 x) + C\sin(\beta_2 x), & (0 < x < t_2) \\ D\cos(\beta_3 x) + E\sin(\beta_3 x), & (-t_3 \leq x \leq 0) \\ [D\cos(\beta_3 t_3) - E\sin(\beta t_3)]e^{\beta_4(x+t_3)}. & (-\infty < x < -t_3) \end{cases} \tag{S6}$$

The magnetic field is found using Ampere's law as $\mathbf{H}_j(x,y) = \frac{1}{Z_0 k_0}\nabla \times \mathbf{E}_j(x,y) = \hat{\mathbf{x}}H_x + \hat{\mathbf{y}}H_y$, with the in-plane component given by:

$$H_y(x) = i\partial_x E_z(x) = \begin{cases} -i\beta_0[D\cos(\beta_2 t_2) + C\sin(\beta_2 t_2)]e^{-\beta_0(x-t_2)}, & (t_2 < x < +\infty) \\ -i\beta_2[D\sin(\beta_2 x) - C\cos(\beta_2 x)], & (0 < x < t_2) \\ -i\beta_3[D\sin(\beta_3 x) - E\cos(\beta_3 x)], & (-t_3 \leq x \leq 0) \\ i\beta_4[D\cos(\beta_3 t_3) - E\sin(\beta_3 t_3)]e^{\beta_4(x+t_3)}. & (-\infty < x < -t_3) \end{cases} \tag{S7}$$

Similarly, the continuity of the magnetic field applied at boundary $x = t_2, 0, -t_3$:

$$\begin{cases} \beta_0[D\cos(\beta_2 t_2) + C\sin(\beta_2 t_2)] = \beta_2(D\sin(\beta_2 t_2) - C\cos(\beta_2 t_2)), \\ \beta_2 C = \beta_3 E, \\ \beta_3[D\sin(\beta_3 t_3) + E\cos(\beta_3 t_3)] = \beta_4[D\cos(\beta_3 t_3) - E\sin(\beta_3 t_3)]. \end{cases} \tag{S8}$$

Combining these equations, we obtain

$$D = -\frac{\beta_3[\beta_2\cos(\beta_2 t_2) + \beta_1\sin(\beta_2 t_2)]}{\beta_2[\beta_1\cos(\beta_2 t_2) - \beta_2\sin(\beta_2 t_2)]}E,$$

$$D = -\frac{\beta_4\sin(\beta_3 t_3) + \beta_3\cos(\beta_3 t_3)}{\beta_3[\sin(\beta_3 t_3) - \beta_4\cos(\beta_3 t_3)]}E, \tag{S9}$$

which yields:

$$\frac{\beta_4\tan(\beta_3 t_3) + \beta_3}{\beta_3\tan(\beta_3 t_3) - \beta_4} = \frac{\beta_3[\beta_2 + \beta_0\tan(\beta_2 t_2)]}{\beta_2[\beta_0 - \beta_2\tan(\beta_2 t_2)]}. \tag{S10}$$

Rearranging, we obtained the general TE dispersion equation that explains the optimized GMR model

$$\tan(\beta_3 t_3) = \frac{\beta_3(\beta_0\beta_4 - \beta_2^2)\tan(\beta_2 t_2) + \beta_3\beta_2(\beta_0 + \beta_4)}{(\beta_3^2\beta_0 + \beta_2^2\beta_4)\tan(\beta_2 t_2) + \beta_2(\beta_3^2 - \beta_0\beta_4)}. \tag{S11}$$

As the thickness of the silicon layer approaches zero, the optimized GMR described in Eq. (S11) degenerates into the dispersion relation of the three-layer conventional GMR model [2]:

$$\tan(\beta_2 t_2) = \frac{\beta_2(\beta_0 + \beta_4)}{\beta_2^2 - \beta_0 \beta_4}. \tag{S12}$$

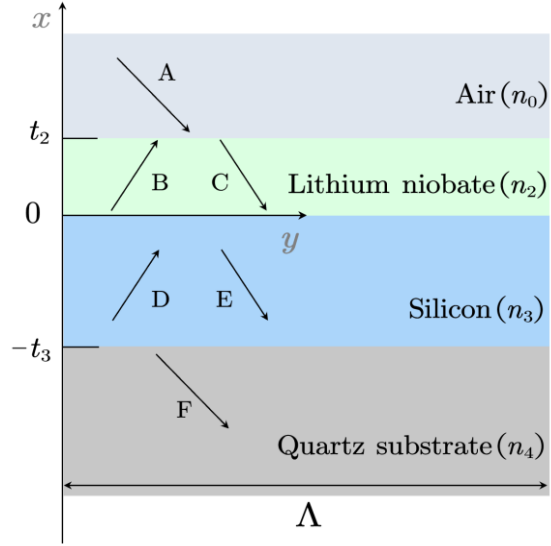

**Fig. S1.** Effective model of proposed GMR structure, consisting of a dual-layer waveguide (non-$\chi^{(2)}$ silicon and $\chi^{(2)}$ lithium niobate), simplified into a four-layer planar structure.

## 2. Supporting Figures

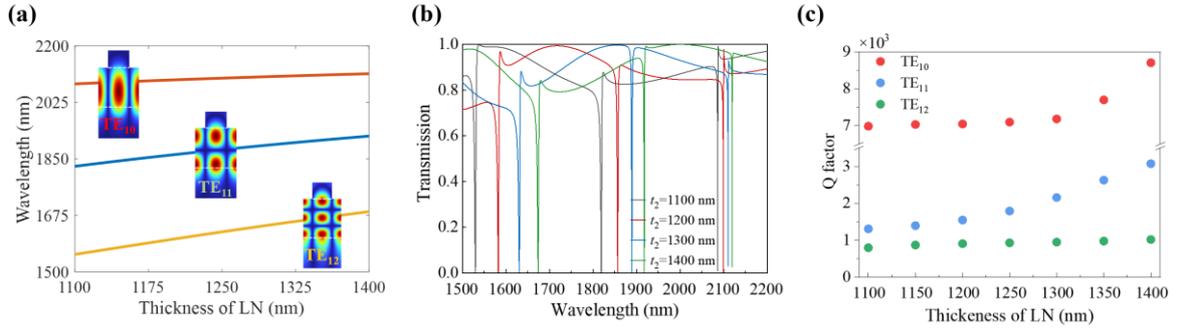

**Fig. S2**. **Linear optical characteristics of a conventional GMR structure with a single-layer LN waveguide configuration** ($t_3 = 0$). (a) Dispersion relations of the three resonant modes as a function of the LN thickness (1100-1400 nm), obtained by solving the transcendental equation Eq. (S12). The inset shows the corresponding electric field eigenmode obtained from COMSOL eigenmode analysis. (b) Mode-resolved spectral responses calculated using rigorous coupled-wave analysis, showing the evolution of the three modes with varying LN thickness (represented by different color). (c) Quality-factors (Q-factors) of the three modes extracted via Fano fitting,[3] indicating strong light–matter interaction in the GMR structure. We focus primarily on the fundamental low-frequency transverse electric $TE_{10}$ mode, which exhibits the highest Q-factor.

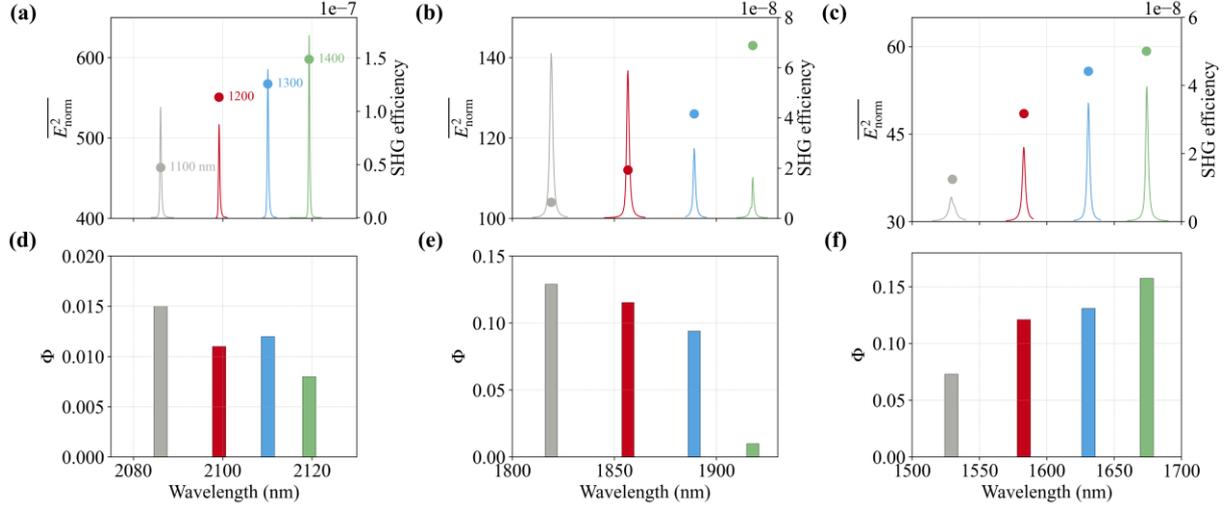

**Fig. S3. Nonlinear analysis of conventional GMR structure.** (a-c) SHG efficiencies (right axis, solid line) and field enhancement factors (left axis, dots) of three modes, $TE_{10}$ (a), $TE_{11}$ (b) and $TE_{12}$ (c), and (d-f) corresponding overall phase-matching factors, are plotted as functions of LN thickness (1100-1400 nm). For the $TE_{10}$ mode, the SHG efficiency (a, right axis, solid line) is determined by both the field enhancement factor and the overall PMF (d). Notably, the abnormal behavior of SHG efficiency, which does not increase monotonically with field enhancement, indicates that the overall PMF plays a critical role—highlighting the importance of phase matching in nanophotonic structures. Panels (b, e) and (c, f) show results for the $TE_{11}$ (middle column) and $TE_{12}$ (rightmost column) modes, respectively, with (b) and (c) depicting their field enhancement and SHG efficiency, and (e) and (f) showing the corresponding overall PMFs. Lower-order modes exhibit stronger field confinement and higher SHG efficiency, whereas their phase-matching factors are generally smaller than those of higher-order modes.

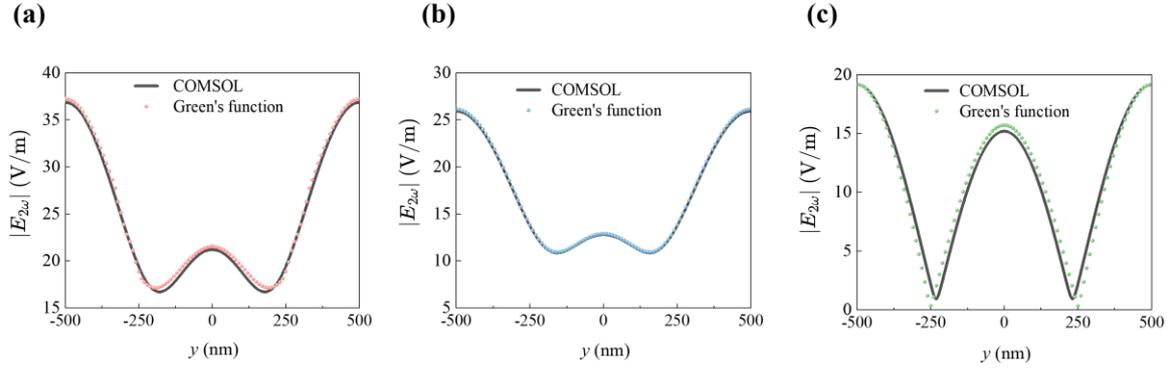

**Fig. S4. Validation of the Green's function method for second-harmonic field calculation.** (a-c) Comparison of the second-harmonic field amplitude of $E_{2\omega}(x,y)$ for TE$_{11}$ mode obtained from the Green's function method (red dot) and finite element simulations (black line) for different LN thicknesses: (a) $t_2$ = 1200 nm, (b) $t_2$ = 1300 nm, and (c) $t_2$ = 1400 nm. The fundamental field distribution $E_\omega(x',y')$ within the lithium niobate layer is recorded through full-wave simulations in COMSOL. The general approach to simulating second harmonic generation in COMSOL Multiphysics, under the undepleted-pump approximation, is to treat the nonlinear polarization $P_{2\omega}(x',y') = \varepsilon_0 \chi_{33}^{(2)} E_\omega^2(x',y')$ at the harmonic wavelength as a source in a separate linear simulation to evaluate the system's nonlinear response. While the second-harmonic field $E_{2\omega}(x,y)$ is computed by Green's function method as follows. First, by sweeping out-of-plane current sources at the harmonic frequency across the entire LN region, we obtained the Green's function $G_{2\omega}(x',y';x,y)$ that characterizes the radiation into the ports from a unit-area polarization located at any position within the nonlinear medium, where the simulated linear field amplitude of the current sources corresponds to the polarization amplitude per unit area $J_{2\omega}(x',y') = -2j\omega P_{2\omega}(x',y')$. Then, the second-harmonic field $E_{2\omega}(x,y)$ is obtained by integrating the product of the Green's function $G_{2\omega}(x',y';x,y)$, which represents the out-of-plane radiation from a current source at $(x',y')$ to the observation point $(x,y)$, and the recorded fundamental field $E_\omega(x',y')$ over the entire lithium niobate region, as described by $E_{2\omega}(x,y) = -\frac{4\omega^2 \chi_{33}^{(2)}}{c^2} \int_{LN} dx'dy' G_{2\omega}(x,y,x',y') E_\omega^2(x',y')$.

The $E_{2\omega}(x,y)$ output here comes from the farthest port as shown in Fig. S5a. The primary source of error arises from spatial misalignment between the field of fundamental wave $E_\omega(x',y')$ and the harmonic Green's function $G_{2\omega}(x',y';x,y)$.

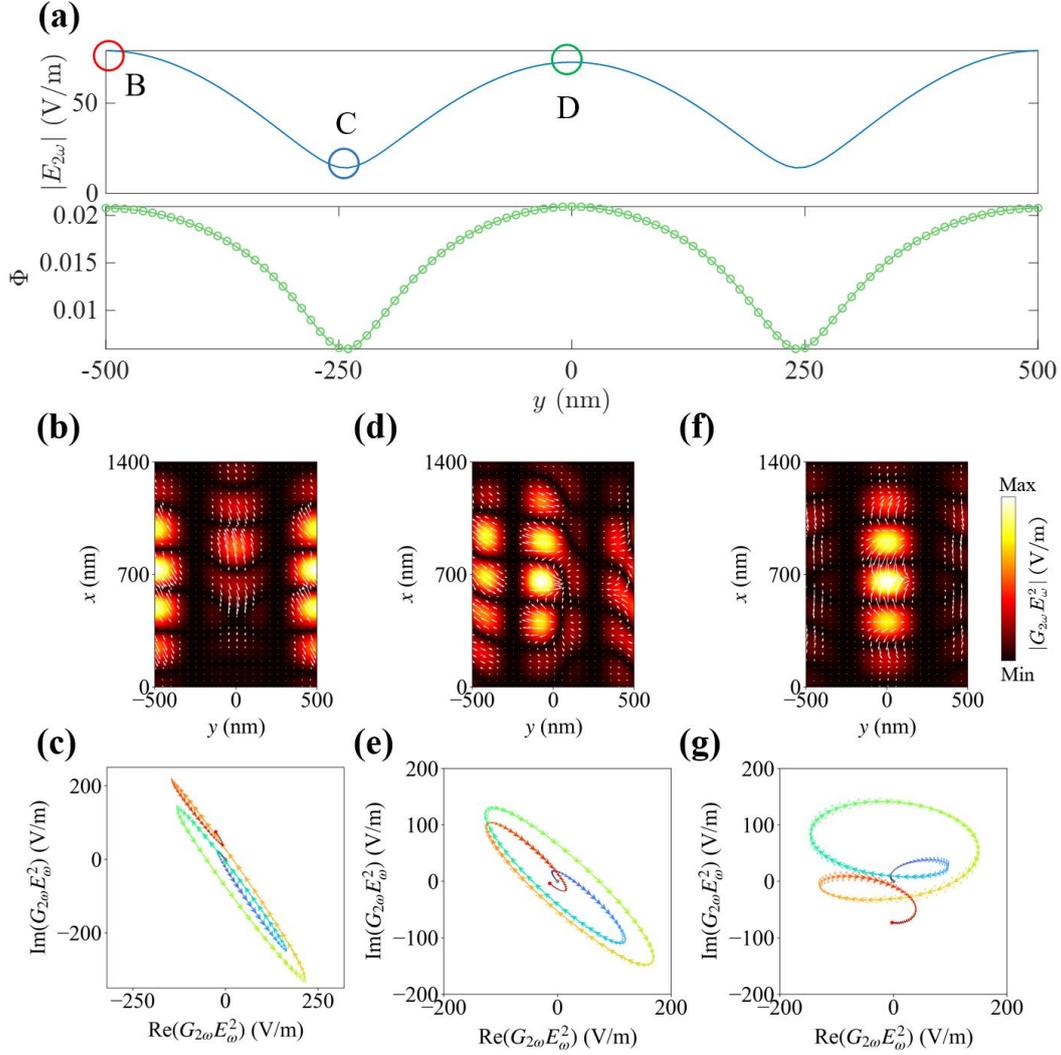

**Fig. S5. Phase-matching analysis of a conventional GMR structure at the representative positions at the output port for the TE$_{10}$ resonant mode.** The LN waveguide thickness $t_2$ = 1400 nm. (a) Harmonic field amplitude (top) and pointwise phase-matching factor (bottom) along the output port in the *y*-direction, with representative positions B, C, and D marked with circles. (b) Amplitude (colormap) and phase (white arrows) distributions of the term $G_{2\omega}E_\omega^2$ contributing to point B. (c) Row-wise vector summation path at point B, where color gradient (blue → red) indicates the sequence of vector addition. The resultant vector (from the origin to the red dot) corresponds to the harmonic field strength of point B in (a). (d, e) Distribution of $G_{2\omega}E_\omega^2$ and interference-based vector summation processes for points C; (f, g) the corresponding results for point D.

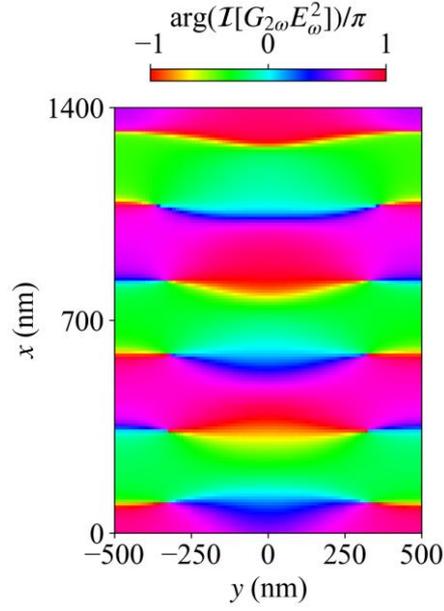

**Fig. S6 Phase distribution of $\mathcal{I}[G_{2\omega}E_\omega^2]$ in a conventional GMR structure.**

Under $TE_{10}$ mode, the LN waveguide thickness $t_2$ = 1400 nm. The phase map suggests constructive interference of the subharmonic contributions along the *y*-direction, but destructive interference along the *x*-direction. This feature ultimately gives rise to the elliptical looping superposition path presented in Fig. 3f.

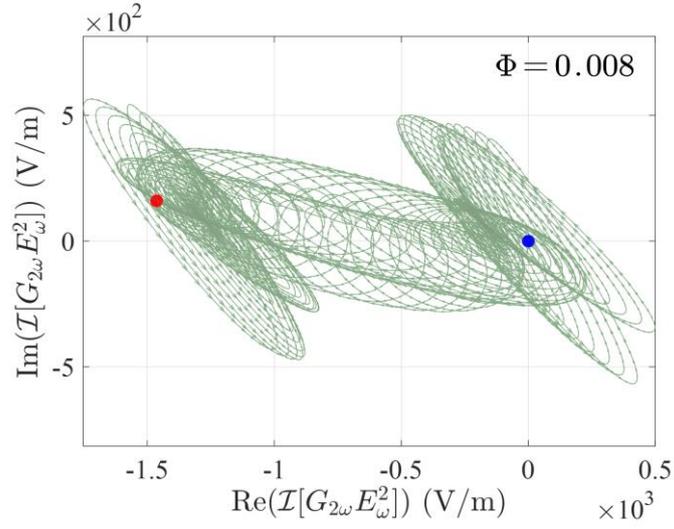

**Fig. S7.** Column-wise summation of the phasor $\mathcal{I}[G_{2\omega}E_\omega^2]$ corresponding to Fig. 3f. Despite the highly irregular phase accumulation pattern, the summation yields the same phase-matching factor and final harmonic field amplitude. The blue and red dots indicate the starting and ending positions of the vector arrows, respectively.

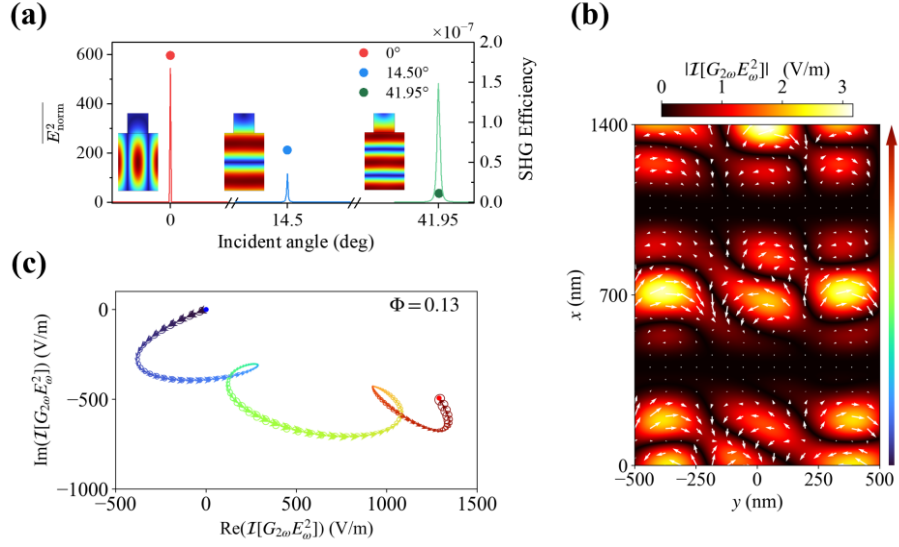

**Fig. S8. Strategy for enhancing phase matching factor via oblique incidence.** (a) Field enhancement factor (left axis, dots) and harmonic generation efficiency (right axis, lines) as functions of incidence angle. At 41.95°, the efficiency is only slightly lower than that under normal incidence. The inset shows the corresponding fundamental field distribution $|E_\omega|$. (b) Spatial distribution of the generalized effective excitation source (proportional to $|\mathcal{I}[G_{2\omega}E_\omega^2]|$) in a nonlinear waveguide at 41.95°, with the white arrows in the background denoting the phaser distribution of $\mathcal{I}[G_{2\omega}E_\omega^2]$. (c) Vectorial superposition of the phasor $\mathcal{I}[G_{2\omega}E_\omega^2]$, yielding a phase-matching factor of 0.13. Although oblique incidence breaks the symmetry of the microscopic nonlinear sources, the interference of subharmonic waves still follows multiple characteristic elliptical superposition trajectory provided by the GFIM.

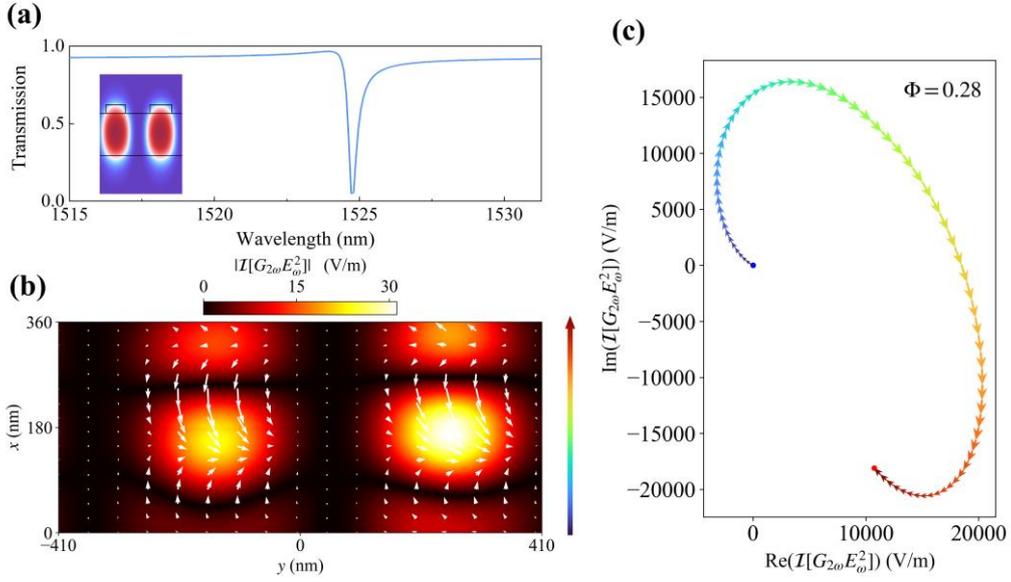

**Fig. S9. Quasi-BIC-based structure with improved phase matching.** (a) Transmission spectrum of a quasi-bound state in the continuum (quasi-BIC) structure reported in Reference [[4]]. The inset shows the corresponding electric field distribution at the resonance wavelength. (b) Spatial distribution of the microscopic nonlinear sources (characterized by $\mathcal{I}[G_{2\omega}E_\omega^2]$) in the LN waveguide. (c) Vectorial superposition of the phasor $\mathcal{I}[G_{2\omega}E_\omega^2]$. A phase-matching factor of 0.28 is achieved, indicating a moderate improvement when the thickness of nonlinear waveguide core is reduced to less than a single interference period.

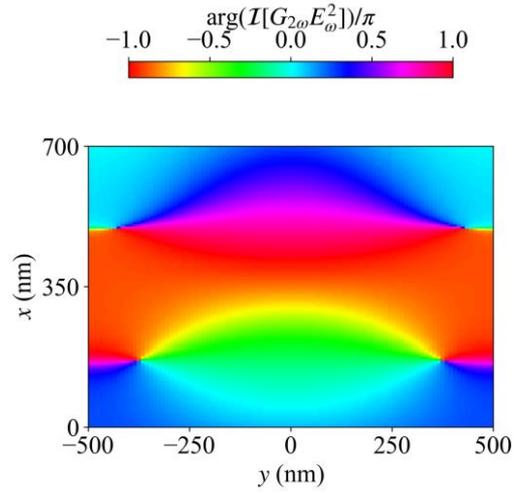

**Fig. S10. Phase distribution of $\mathcal{I}[G_{2\omega}E_\omega^2]$ in optimized dual-layer GMR structure under TE$_{10}$ mode pumping.** The slowly varying phase in the regions of strong fundamental field enhancement markedly promotes the increase of the phase-matching factor.

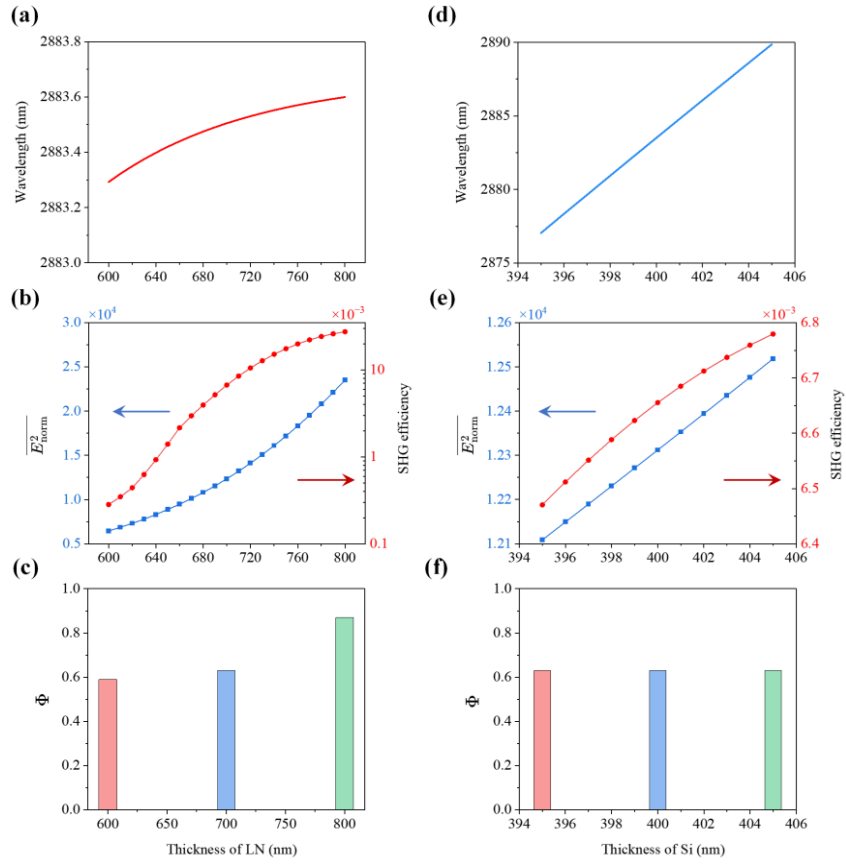

**Fig. S11 Thickness tuning of a bilayer waveguide for enhanced nonlinear efficiency.** (a) Dispersion relation as a function of LN thickness $t_2$ ($t_3 = 400$ nm). (b) Variation of the field enhancement factor $\overline{E_{\text{norm}}^2}$ (left axis, blue line) and SHG efficiency (right axis, red line) with LN thickness. The SHG efficiency reached 2.75% at 800nm-thick lithium niobate. (c) Phase-matching factor corresponding to $t_2 = 600, 700, 800$ nm. (d) Dispersion relation as a function of silicon thickness $t_3$ ($t_2 = 700$ nm). (e) Variation of the field enhancement factor $\overline{E_{\text{norm}}^2}$ (left axis, blue) and SHG efficiency (right axis, red) with silicon thickness. (f) Phase-matching factor corresponding to $t_3 = 395, 400, 405$ nm. Under these favorable phase-matching conditions, increasing the thickness of either the LN or Si waveguide layer leads to a continuous rise in $\overline{E_{\text{norm}}^2}$, which remains on the order of $10^4$, providing a practical approach to achieving high SHG efficiency through thickness tuning. The SHG efficiency for 405nm-thick silicon layer is $6.8\times10^{-3}$.

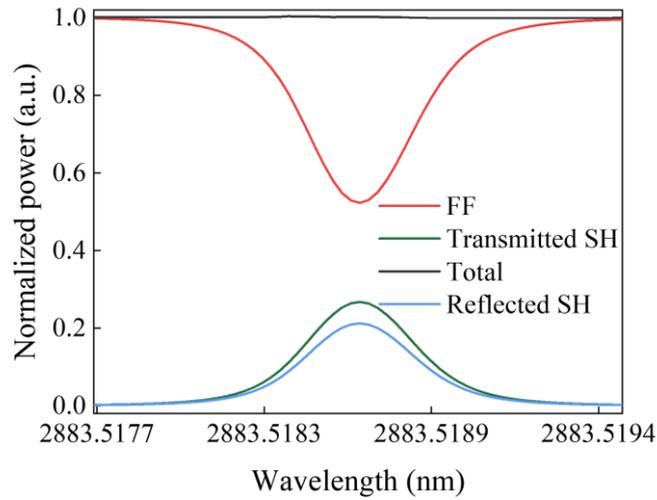

**Fig. S12. Conversion efficiency obtained from a coupled simulation model of fundamental and second-harmonic waves.** This simulation is applicable to the depleted-pump regime, as validated by the overall energy conservation. The power of fundamental frequency (FF, red) and second harmonic (SH) from both the transmission (green) and reflection (blue) ports are considered.

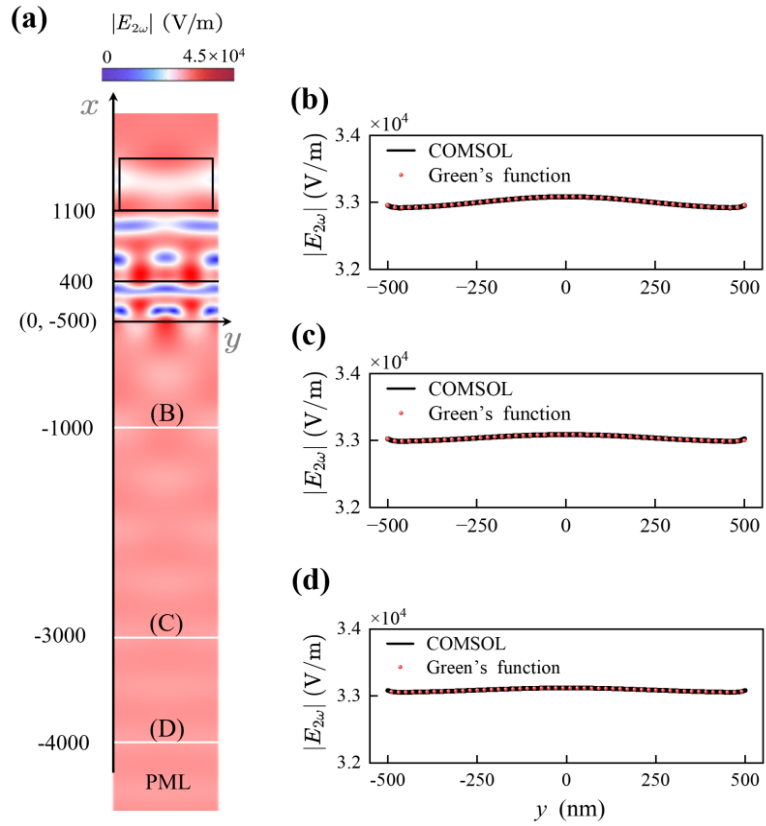

**Fig. S13 Harmonic field distributions within the optimized GMR structure for a PMMA width of 900 nm.** (a) Spatial distribution of the harmonic field at a power density of 2 kW/cm², with the color bar indicating the maximum and minimum values. (b-d) Magnitudes of the harmonic fields at the three corresponding labeled ports. The observed spatial uniformity indicates an approximately plane-wave radiation.